\documentclass[epj]{svjour}
\usepackage{graphics}
\usepackage{bm}
\usepackage{epsfig}
\usepackage{graphicx}
\usepackage{amsfonts}
\usepackage{float}
\usepackage{amsmath,amssymb}

\newcommand{\nn}{\nonumber}
\newcommand{\beq}{\begin{equation}}
\newcommand{\eeq}{\end{equation}}
\newcommand{\bea}{\begin{eqnarray}}
\newcommand{\eea}{\end{eqnarray}}
\newcommand{\bp}{{\bm p}}
\newcommand{\bP}{{\bm P}}
\newcommand{\bq}{{\bm q}}
\newcommand{\bx}{{\bm x}}
\newcommand{\br}{{\bm r}}

\newcommand{\bsigma}{{\bm \sigma}}
\newcommand{\brho}{{\bm \rho}}
\newcommand{\blambda}{{\bm \lambda}}

\begin{document}
\title{SU(4) flavor symmetry breaking in D-meson couplings to light hadrons}
\author{C.E. Fontoura\inst{1,2}
\and J. Haidenbauer\inst{3} 
\and G. Krein \inst{2}}               
\institute{Instituto Tecnol\'ogico da Aeron\'autica, DCTA, 12228-900 S\~ao Jos\'e dos Campos, SP, Brazil
\and
Instituto de F\'{\i}sica Te\'orica, Universidade Estadual Paulista, 01140-070 S\~ao Paulo, SP, Brazil
\and 
Institute for Advanced Simulation, Institut f\"ur Kernphysik, and J\"ulich Center for Hadron Physics \\
Forschungszentrum J\"ulich, D-52425 J\"ulich, Germany
}
\date{Received: date / Revised version: date}
%
\abstract{
The validity of SU(4)-flavor symmetry relations of couplings of charmed $D$ mesons to light
mesons and baryons is examined with the use of $^3{\rm P}_0$ quark-pair creation model and nonrelativistic 
quark model wave functions. We focus on the three-meson couplings $\pi\pi\rho$, $KK\rho$ and
$DD\rho$ and baryon-baryon-meson couplings $NN\pi$, $N\Lambda K$ and $N\Lambda_c D$. It is 
found that SU(4)-flavor symmetry is broken at the level of 30\% in the $DD\rho$ tree-meson
couplings and 20\% in the baryon-baryon-meson couplings. Consequences of these findings for 
DN cross sections and existence of bound states D-mesons in nuclei are discussed. 
} 
\maketitle
%


\noindent{\it Introduction.} Currently there is considerable interest in exploring the interactions 
of charmed hadrons with light hadrons and atomic nuclei~\cite{Briceno:2015rlt}. Particular attention 
is paid to $D$ mesons, much discussed over the last few years in connection with $D$-mesic 
nuclei~\cite{{Tsushima:1998ru},{GarciaRecio:2010vt},{GarciaRecio:2011xt}} and $J/\psi$ binding 
to nuclei~\cite{{Ko:2000jx},{Krein:2010vp}}. Presently, there is no experimental 
information about the $DN$ interaction, a situation that the $\overline{\rm P}$ANDA@FAIR 
ex\-pe\-r\-i\-ment \cite{Wiedner:2011mf} could remedy in the future. Most of the knowledge on the
$DN$ interaction comes from calculations using hadronic Lagrangians motivated by SU(4) extensions 
of light-flavor chiral Lagrangians~\cite{{Mizutani:2006vq},{Lin:1999ve},{Hofmann:2005sw},{Haidenbauer:2007jq},{Haidenbauer:2008ff},{Haidenbauer:2010ch},{Fontoura:2012mz}} and heavy quark 
symmetry~\cite{{Yasui:2009bz},{GarciaRecio:2008dp}}. These require as input coupling constants and, 
in some cases, form factors. For the particular case of $\bar{D}N$ reactions (where 
$\bar D \equiv \{\bar{D}^0, D^-\}$), Ref.~\cite{Haidenbauer:2007jq} found that among all the couplings 
in the ef\-fec\-tive Lagrangian, $g_{DD\rho}$ and 
$g_{DD\omega}$ 
provide the largest 
contributions to cross sections and phase shifts for kinetic center of mass (c.m.) energies up to $150$~MeV 
{\textemdash} they also play an important role for the $DN$ interaction~\cite{Haidenbauer:2010ch}. 
Flavor SU(4) symmetry relates those couplings to couplings in the light-flavor sector:  
\bea
&& g_{DD\rho} = g_{KK\rho} = \frac{1}{2} g_{\pi\pi\rho}, 
\label{mes-SU4} \\
&& g_{N\Lambda_{c} D} = g_{N\Lambda K} = \frac{3\sqrt{3}}{5} g_{NN\pi}.
\label{bar-SU4}
\eea
The studies in Refs~\cite{Haidenbauer:2007jq,Haidenbauer:2008ff,Haidenbauer:2010ch} utilized
the SU(4) relations above, based on $g_{\pi\pi\rho} = 6.0$ and $g_{NN\pi} = 13.6$, which 
are the values used in a large body of work conducted within the J\"ulich 
model~\cite{{Haidenbauer:1991kt},{Hoffmann:1995ie}} for light-flavor hadrons. 

Given the prominent role played by meson-baryon Lagrangians in the study of the $DN$ interaction and associated 
phenomena, it is of utmost importance to assess the validity of (\ref{mes-SU4}) and (\ref{bar-SU4}). 
SU(4) breaking effects on three-hadron couplings were examined recently using a variety of approaches, 
that include vector meson dominance (VMS)~\cite{Mat98,Lin00a}, Dyson-Schwinger and Bethe-Salpeter equations 
(DS-BS) of QCD~\cite{ElBennich:2011py}, QCD sum rules (QCDSR)~\cite{{Bracco:2011pg},{lc-sr},{qcdsr}}, 
lattice QCD~\cite{Can:2012tx}, and holographic QCD~\cite{Ballon-Bayona:2017bwk}. 
In this work we use the quark model with a $^3{\rm P}_0$ quark-pair creation operator~\cite{3P0}. 
In this setting, the three-hadron couplings are given by matrix elements of the $^3{\rm P}_0$ operator 
evaluated with quark-model wave functions. The literature on the $^3{\rm P}_0$ model is too vast to 
be properly reviewed here, we simply mention that it is being used extensively since the early 1970s 
to study strong decays and that our calculation of vertices shares similarities with those 
of nucleon-meson couplings and form factors~in~\cite{{3P0},{Downum:2006re}}.

\vspace{0.2cm}
\noindent
{\it Three-Hadron Couplings.} To evaluate the matrix element of the $^3{\rm P}_0$  quark-pair creation 
operator, $\hat O_{\rm pc}$, it is convenient to employ the ``decay frame" of an initial hadron at 
rest~\cite{3P0,Downum:2006re}, i.e. the transition of a hadron state $|h_1\rangle$ into a final 
two-hadron state $|h_2 h_3\rangle$ is written as
\begin{equation}
\langle h_2h_3|\hat O_{\rm pc}|h_1\rangle \equiv \delta(\bP_1 - \bP_2 - \bP_3) \, 
{\cal M}_{h_1h_2h_3}(\bq),
\label{trans}
\end{equation}
with $\bq = \bP_2 = - \bP_3$, and
\beq
\hat O_{\rm pc} = \gamma \sum_{cfss'}\int d^{3}p \; 
\bsigma^{c}_{s's}\cdot \bp \;
q^{cf\dag}_{s'}(\bp)\bar{q}^{cf\dag}_{s}(-\bp),
\label{Opc}
\eeq
where $\gamma$ gives the strength of the quark-pair creation, 
$q^{cf\dag}_{s'}(\bm{p})$ and  $\bar{q}^{cf\dag}_{s}(\bm{p})$ are creation operators with
color $c$, flavor $f$, spin projection $s$, and momentum ${\bp}$, $\bm{\sigma}^{c}_{s's}=\chi^{\dagger}_{s'}
\bm{\sigma}\chi^{c}_{s}$, with ${\bm \sigma} = (\sigma^1,\sigma^2,\sigma^3)$ being the 
Pauli matrices, $\chi_s$ Pauli spinors, and $\chi^c_s = -i\,\sigma^2\chi^\ast_s$.

We employ the standard quark-model Hamiltonian~\cite{Swanson:1992ec}:
\begin{eqnarray}
H &=& \sum_{i} \left(m_{i}+\frac{p^{2}_{i}}{2m_{i}}\right)
- \sum_{i<j} \bm{F}_{i}\cdot\bm{F}_{j} 
\biggl[\left(\,\frac{3}{4}b\,r_{ij} - \frac{\alpha_{c}}{r_{ij}}\right) \nn \\
&& + \, \frac{8\,\pi\,\alpha_{s}}{3\,m_{i}\,m_{j}} 
\left(\,\frac{\sigma^{3}}{\pi^{\frac{3}{2}}}\,e^{-\sigma^{2}r_{ij}^{2}}\,\right)
{\bm S}_{i}\cdot {\bm S}_{j}\biggr],
\label{QM-H}
\end{eqnarray}
where $m_i$ are the quark masses and ${\bm F}={\bm \lambda}/2$, with ${\bm \lambda}$ the color
SU(3) Gell-Mann matrices and ${\bm S}$ the spin-1/2 vector. Notwithstanding the inability of the model 
to describe all features associated with the Goldstone-boson nature of the pion, nonetheless it 
mimics some of the effects of dynamical chiral symmetry breaking, notably the $\pi-\rho$ mass 
splitting~\cite{Szczepaniak:2000bi}. As in QCD itself, the only source of SU(4) 
breaking in (\ref{QM-H}) is the quark-mass matrix and hence the breaking in the couplings comes solely 
from the hadron wave functions. The Schr\"odinger equation is solved as a generalized matrix problem 
using a finite basis of Gaussian functions with the eigenvalues determined by the Rayleigh-Ritz 
variational principle. Reasonable values for the masses of the 
ground states of the hadrons of interest can be obtained by expanding the meson $\Phi$ and 
baryon $\Psi$ intrinsic wave functions as~\cite{{Swanson:1992ec},{SilvestreBrac:1995gz}}:
\bea
\hspace{-0.3cm}\Phi(\br) = \sum_{n=1}^{N} c_{n}\, \varphi_n(\br), 
\hspace{0.20cm}\Psi(\brho,\blambda) = \sum_{n=1}^{N} c_{n}\, \varphi_n(\brho)\varphi_n(\blambda),
\eea
where the $c_n$ are dimensionless expansion parameters and
\beq
\varphi_n(\bx) = \left(\frac{n\alpha^2}{\pi}\right)^{3/4}
\, e^{- n \alpha^2 \bx^2/2}.
\eeq
Here, $\alpha$ is the variational, $\br = \br_1 - \br_2$, $\brho = (\br_1 - \br_2)/\sqrt{2}$,
and $\blambda= \sqrt{2/3}[(\br_1+\br_2)/2 - \br_3]$. The matrix element ${\cal M}(\bq)$ can be evaluated
analytically; it is given by 
\beq
{\cal M}_{h_1h_2h_3}(\bq) = \kappa_{h_1h_2h_3} \, A_{h_1h_2h_3}(\bq) \, |\bq| \, Y_{1m}(\hat\bq) ,
\label{M-gen}
\eeq
where $Y_{1m}(\hat\bq)$ are spherical harmonics with $m=1 (0)$ for three-meson 
(nucleon-baryon-meson) couplings, $\kappa$ comes from summing over color, spin, and flavor 
and is given by 
\bea
&&\kappa_{DD\rho} = \kappa_{KK\rho} =  \frac{1}{2}\kappa_{\pi\pi\rho} = 1, \\ 
&&\kappa_{N\Lambda_c D} = \kappa_{N\Lambda K} =  \frac{3\sqrt{3}}{5} \kappa_{NN\pi} = 1.  
\eea
The am\-pli\-tude $A_{h_1h_2h_3}(\bq)$ in (\ref{M-gen}) is given by 
\bea
&&A_{h_1h_2h_3}(\bq) =  \gamma \, \sum_{n_1n_2n_3} c^*_{n_3} c^*_{n_2} c_{n_1}  
(n_1n_2n_3)^{3/4} 
\nonumber \\ 
&& \hspace{0.75cm}\times \, f_{h_1h_2h_3}(n_1,n_2,n_3) \,  e^{-\bq^2/\Lambda^2_{h_1h_2h_3}(n_1,n_2,n_3)} ,
\label{A-ampl}
\eea
where $f_{h_1 h_2 h_3}$ are given by ($P$ in $PP\rho$ stands for $\pi,K,D$
and $B$ in NBP for $N,\Lambda,\Lambda_c$)
\bea
f_{PP\rho}(n_1,n_2,n_3) &=& \left(\frac{64}{9\pi}\right)^{1/4} \frac{\alpha^{3/2}_\rho}{\alpha^3_P}
\nonumber \\
&&\hspace{-2.5cm}\times \,  
\frac{ n_1 n_2  + \left(\overline m_1 \, n_1 n_3 + 2 \, n_2 n_3 \right) \, \alpha^2_\rho/\alpha^2_P }
{\left[n_1 n_2 + (n_1 + n_2) n_3 \, \alpha^2_\rho/\alpha^2_P \right]^{5/2}} , \\
f_{NBP}(n_1,n_2,n_3) &=& \frac{72}{\pi^{3/4}} 
\frac{\alpha^3_B}{\alpha^3_N \alpha^{3/2}_P}
\frac{1}{(n_1 + n_2 \, \alpha^2_B/\alpha^2_N)^{3/2}} \nn \\
&&\hspace{-2.5cm} \times \,
\frac{ {\overline m}_2  n_1 n_2 \, \alpha^2_B/\alpha^2_P + {\widetilde m}_2 n_1 n_3 
+ 3 n_2 n_3 \, \alpha^2_B/\alpha^2_N}
{\left(2 n_1 n_2 \,\alpha^2_B/\alpha^2_P  + 3 n_1 n_3 + 3 n_2 n_3\alpha^2_B/\alpha^2_N
\right)^{5/2}},
\eea
and the ``cut-off" parameters $\Lambda_{h_1 h_2 h_3}$ are given by
\bea
\hspace{-0.55cm}
\Lambda^2_{PP\rho}(n_1,n_2,n_3) &=& \frac{8 \alpha^2_P\!\left[n_1 n_2 + (n_1 + n_2) n_3 \alpha^2_\rho/\alpha^2_P
\right] }{(\Delta\overline m)^2 \, n_1 + n_2 + n_3 \, \overline{m}^2_h \, \alpha^2_\rho/\alpha^2_P }, \\
\hspace{-0.55cm}\Lambda^2_{NBP}(n_1,n_2,n_3) &=& \frac{24 \alpha^2_P}{{\overline m}^2_1 } 
\nonumber \\
&&\hspace{-2.5cm} \times \,
\frac{ 2 n_1 n_2 \alpha^2_B/ \alpha^2_P + 3 n_1 n_3 
+ 3 n_2 n_3 \alpha^2_B/\alpha^2_N }
{n_1 {\widetilde m}^2_2    + 9 n_2 \, \alpha^2_B/\alpha^2_N 
+ 6 ({\widetilde m}_1 + {\widetilde m}_2)^2\,n_3 \alpha^2_P/\alpha^2_N}, 
\eea
where
\beq
{\overline m}_{1,2} = \frac{2 m_{1,2}}{m_2 + m_1},\hspace{0.2cm}
{\widetilde m}_{1,2} = \frac{3 m_{1,2}}{2m_1 + m_2},\hspace{0.2cm}
\Delta \overline m = \frac{m_2-m_1}{m_2+m_1},
\label{bar-m}
\eeq
with $m_1 = m_u =m_d$, $m_2 = m_s, m_c$. 

In the limit of SU(4) symmetry, $m_1 = m_2$, $\alpha_D = \alpha_K = \alpha_\pi$ 
and $\alpha_{\Lambda_c} = \alpha_\Lambda = \alpha_N$, and the ratios 
\beq
{\cal R}_{P/P'}(\bq^2) = \frac{A_{PP\rho}(\bq^2)}{A_{P'P'\rho}(\bq^2) },\hspace{0.125cm} 
{\cal R}_{BP/B'P'}(\bq^2) = \frac{A_{NBP}(\bq^2) }{A_{NB'P'}(\bq^2)},
\label{ratios}
\eeq
are all equal to~1, expressing the same symmetry as in (\ref{mes-SU4}) and~(\ref{bar-SU4}).
In this limit, $\gamma$ must be the same for all couplings, which seems a reasonable assumption, 
as they involve the same light-quark pair creation. Symmetry-breaking effects are contained 
in the factors $f$, $c_{n}$ and $\Lambda$. 

Let us now connect to meson-exchange models. A typical three-meson vertex function, as it
appears in that approach in the $PN$ potentials (with $P=K,\, \bar K,\, \bar D,\, D$)
\cite{{Haidenbauer:2007jq},{Haidenbauer:2008ff},{Haidenbauer:2010ch}},
is given by (in the decay frame)
\beq
A_{PPV}(\bq^2) = \phi_{\rm KF}  \;  g_{PPV} \left(\frac{\Lambda^2_{PPV} - m^2_V}
{\Lambda^2_{PPV} - q^2}\right)^{n} |\bq| Y_{11}(\hat\bq).
\label{A-mex}
\eeq
Here $\phi_{\rm KF}$ is a kinematical factor involving the energies of the hadrons, 
$g_{PPV}$ is the Lagrangian coupling constant, and there is also form factor with a 
cutoff mass $\Lambda_{PPV}$, where $n=1$ or $n=2$~\cite{{Haidenbauer:1991kt},{Hoffmann:1995ie}}. 
Here, the value of $g_{PPV}$ refers to the case when the vector meson $V$ 
is on its mass shell. Then $q^2 = (q^0)^2 - \bq^2 = m^2_V$ and the form factor~is~$1$. 
For low-energy elastic $PN$ scattering, the exchanged $\rho$ (and $\omega$) meson 
is far from its mass shell; the momentum transfer $q^2$ is small and negative, 
i.e. $q^2 = (q^0)^2 - \bq^2 \equiv -\bq^2$ with $\bq^2\gtrsim 0$. Therefore, it is common 
practice to use the static approximation $q^2 = -\bq^2$ in the form factors. 
We note that for the $DN$ ($\bar DN$) processes studied in 
Refs.~\cite{{Haidenbauer:2007jq},{Haidenbauer:2008ff},{Haidenbauer:2010ch},{Fontoura:2012mz}} 
up to kinetic c.m. energy of $150$~MeV, the highest c.m. momentum is $400~{\rm MeV}/c$. 
The cutoff mass in the form factors is another source of symmetry breaking 
in the meson-exchange potentials. However, in the $DN$ ($\bar DN$) interactions in 
\cite{{Haidenbauer:2007jq},{Haidenbauer:2008ff},{Haidenbauer:2010ch}} those masses were
simply taken over from the corresponding $\bar KN$ ($KN$) interactions, for $\rho$- as well as
for $\omega$ exchange. Thus, they drop out in the ratio~(\ref{ratios}). 

The situation with baryon exchange is much more complicated, as different baryons are exchanged 
in the $\bar DN$ and $DN$ reactions. The separation of kinematical effects and the coupling 
strength, as in (\ref{A-mex}), cannot be easily done. Indeed in $\bar DN$ ($KN$) elastic scattering 
only $B=\Lambda_c$ ($\Lambda$) exchange contributes while for $DN$ ($\bar KN$) there is only $N$
exchange, and only in the transitions $DN \leftrightarrow \pi \Lambda_c$ 
($\bar KN \leftrightarrow \pi \Lambda$). Furthermore, for heavy baryons like $\Lambda_c$ an 
extrapolation to the pole is rather 
questionable as the quark-model is not expected to work at such high momenta. 
Despite these drawbacks, we include here our baryon results for illustration purposes.

\vspace{0.2cm}
\noindent{\it Results.} We use the quark-model parameters of~\cite{Swanson:1992ec}:
$m_{l}= 375~{\rm MeV}$, $m_{s}=650~{\rm MeV}$, $\alpha_{c} = 0.857$, $\alpha_s = 0.84$,
$b=0.154~\text{GeV}^{2}$, $\sigma=70~\text{MeV}$. We take $m_c = 1657~{\rm MeV}$ to 
fit the $D$ meson mass. Tab.~\ref{tab1} shows the results; convergence is achieved with 
$N=11$ Gaussian functions. 
Clearly, the model fits well the experimental values of the masses, the largest discrepancy 
is~$4$\% in the mass of $\Lambda_c$. In particular, the $\rho-\pi$ and $N-\Lambda$ mass splittings 
are well described. In addition, $m_\Sigma-m_\Lambda=82$~MeV and $m_{\Sigma_c}-m_{\Lambda_c}=135$~MeV,
also in fair agreement with data~\cite{PDG}. Since the corresponding effects on the $\Sigma$ and $\Sigma_c$ 
wave functions have a very small effect on the coupling constants, we consider only those couplings 
involving $\Lambda$ and $\Lambda_c$. We take $m_u=m_d$ and so $m_\rho = m_\omega$.

\vspace{-0.5cm}
\begin{table}[h]
\caption{Calculated hadron masses ($m_{\rm calc}$) and sizes ($\alpha$). Experimental 
values for the masses ($m_{\rm exp}$) are from the PDG~\cite{PDG}. 
All values are in MeV.}
\label{tab1}  
\begin{center}
\begin{tabular}{ccccc|ccc}
\hline\noalign{\smallskip}
{      } & $\pi$ & $K$ & $D$ & $\rho$ & $N$ & $\Lambda$ & $\Lambda_c$ \\ 
\noalign{\smallskip}\hline\noalign{\smallskip}
$m_{\rm calc}$ & 138 & 495 & 1866 & 770 & 958 & 1115 & 2195 \\
$m_{\rm exp}$  & 138 & 495 & 1866 & 770 & 940 & 1115 & 2286 \\
\noalign{\smallskip}\hline\noalign{\smallskip}
$\alpha$       & 359 & 377 &  499 & 275 & 234 &  241 &  253 \\
\noalign{\smallskip}\hline
\end{tabular}
\end{center}
\end{table}

\begin{figure}[t]
\begin{center}
\resizebox{9.cm}{!}{\includegraphics{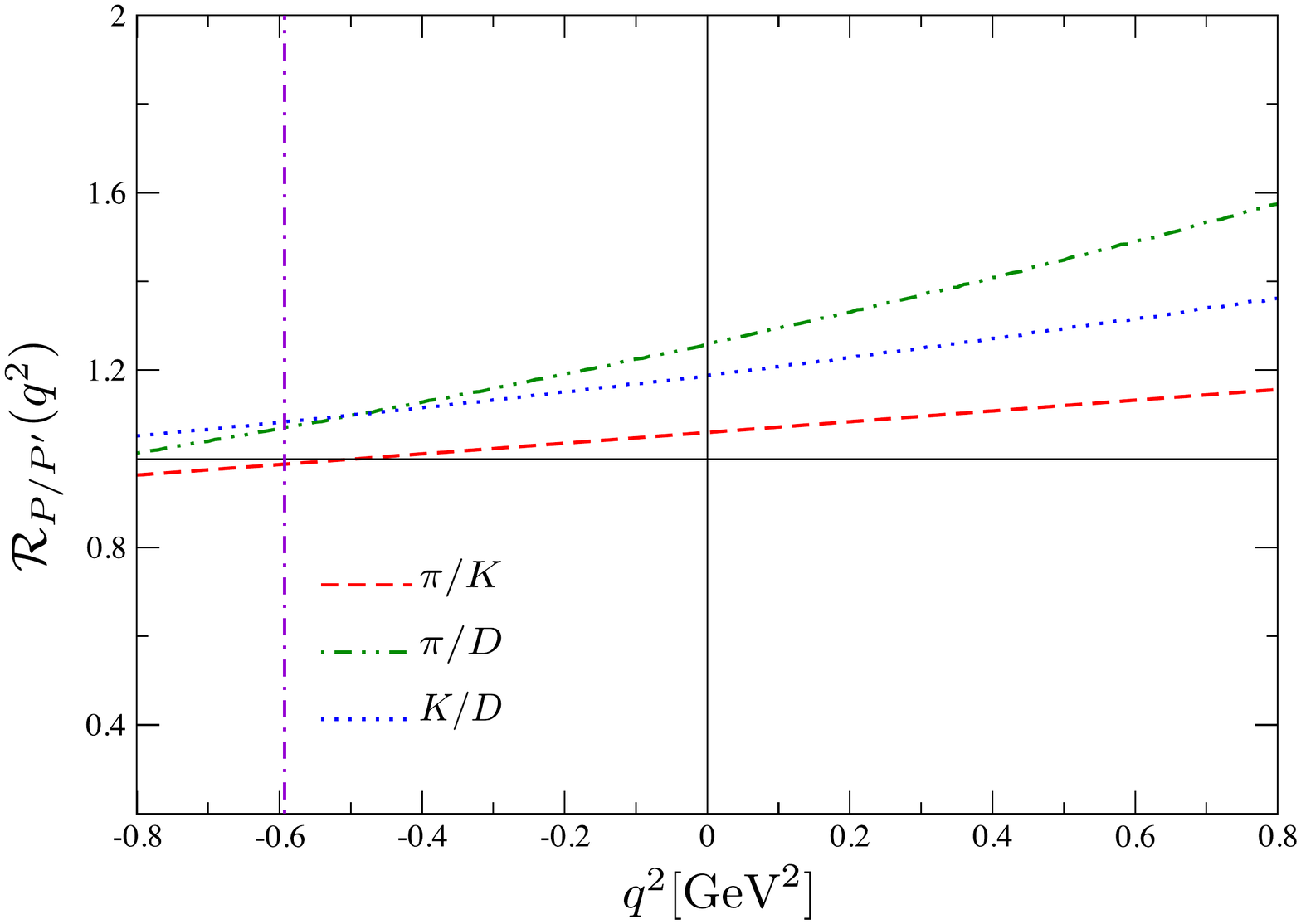}} \\[-0.4true cm]
\resizebox{9.cm}{!}{\includegraphics{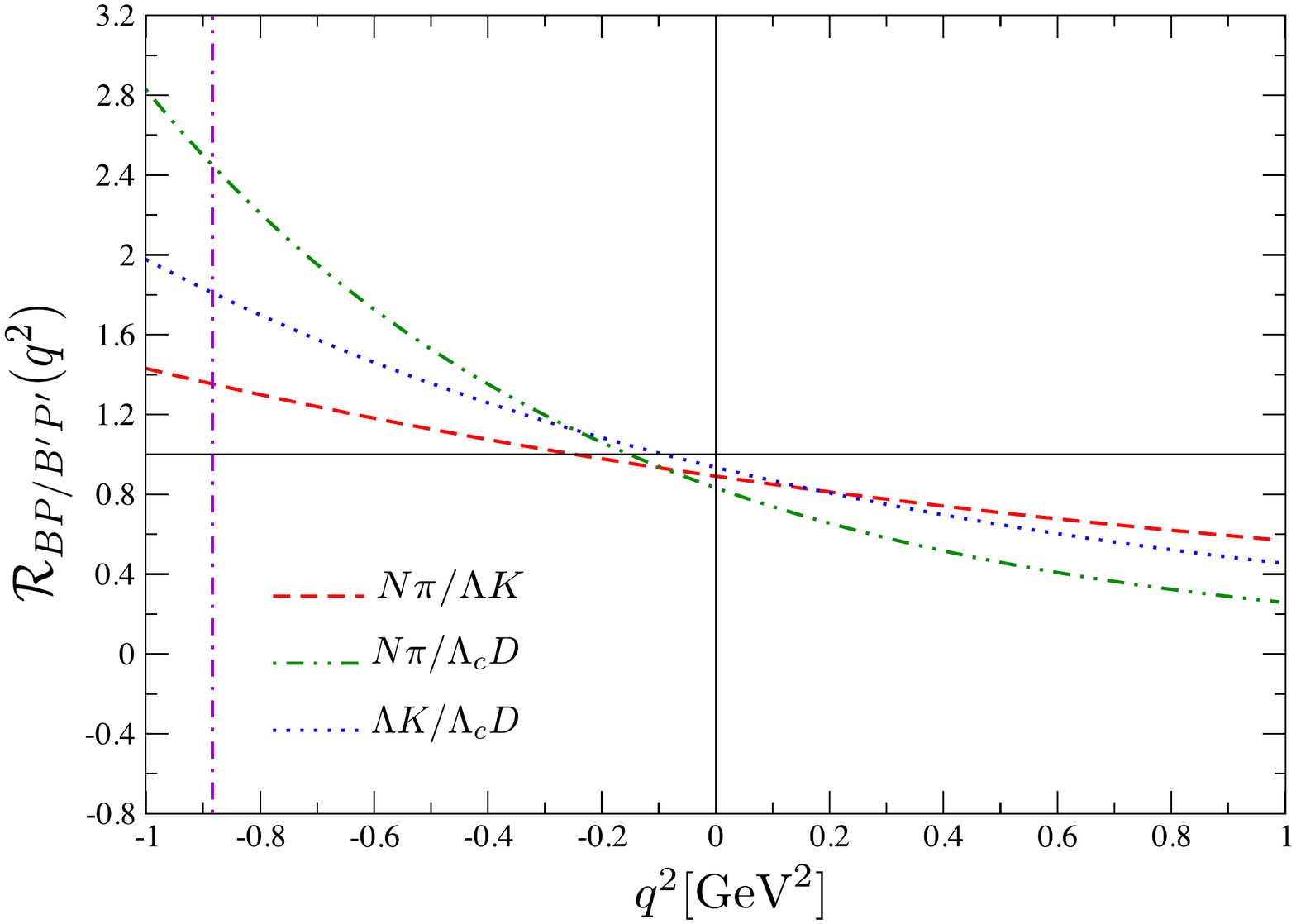}} 
\caption{Ratios ${\cal R}_{P/P'}$ and ${\cal R}_{BP/B'P'}$; the vertical lines
correspond to $\bq^2 = - m^2_\rho$ (top) and $\bq^2 = - m^2_N$ (bottom).}
\label{fig1}
\end{center}
\end{figure}

\vspace{-0.5cm}
The ratios ${\cal R}(\bq^2)$ are shown in Fig.~\ref{fig1}; we recall, $PP\rho$ couplings 
enter graphs with $\rho$ exchange and $NBP$ couplings in graphs with baryon $B$ exchanges. 
Fig.~\ref{fig1} reveals that SU(4) breaking, at $\bq^2 = 0$ and $\bq^2 = - m^2_\rho$, 
is relatively modest. At $\bq^2=0$, 
the largest SU(4) breaking, not unexpectedly, is in $DD\rho$, of the order of $30$\% compared
to $\pi\pi\rho$ coupling, and $20$\% compared to $KK\rho$. Moreover, in agreement with phenomenology, 
there is almost no SU(3) breaking in $KK\rho$. At the $\rho$ pole ($\bq^2 = - m^2_\rho$) the breaking 
is also small, at most $10$\% in $DD\rho$ coupling. The ratios of $NBP$ couplings are presented in the bottom 
panel of the figure. As can be seen, the SU(4) breaking at $\bq^2~=~0$ is at most $20$\% in the $N\Lambda_cD$ 
vertex compared to the $NN\pi$ coupling and $10$\% compared to the $N\Lambda K$.  
The SU(3) symmetry breaking, i.e. in the $N\Lambda K$ coupling, is of the order of $10$\%, also compatible 
with phenomenology. 
Interestingly, for $\bq^2 \approx - 0.9$ GeV/c, i.e close to the nucleon pole (for orientation,
shown by the vertical line in the bottom panel of Fig.~\ref{fig1}), the $N\Lambda_cD$ 
coupling is 3~times smaller than the $NN\pi$ coupling, while the ratio of the $N\Lambda K$ to $N\Lambda_cD$ 
couplings is around 1.8. This is to be compared with the value~0.68 in~\cite{lc-sr}. However, such possible 
SU(4) breaking far into the time-like region might not be relevant for low-energy $\bar DN$ 
scattering because, according to~\cite{Haidenbauer:2007jq}, the contribution of $\Lambda_c$ 
exchange to the $\bar DN$ cross section is very small anyway. 

\begin{figure}[h]
\begin{center}
\resizebox{9.cm}{!}{\includegraphics{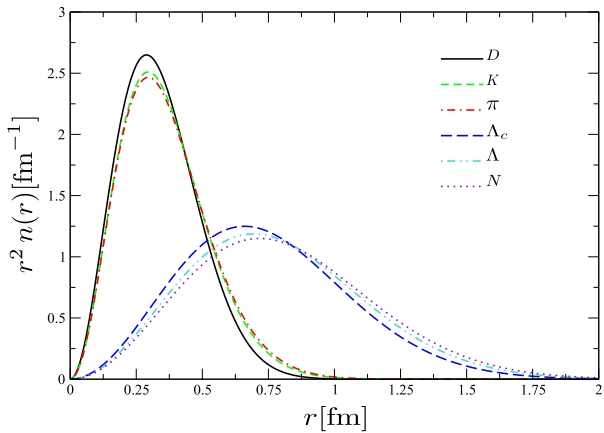}}
\vspace{-0.5cm}
\caption{Normalized light-quark radial distributions in mesons and baryons.}
\label{fig2}
\end{center}
\end{figure}

Physically, the SU(4) breaking originates from the different extensions of the hadron wave functions.
In Fig.~\ref{fig2}, we plotted the normalized light-quark radial distribution functions in the hadrons of 
interest{\textemdash}the Fourier transform of $\langle h| q^\dag(\bq)q(\bq)|h\rangle$. 
The distributions get more compact (shorter-ranged) for heavier hadrons as the binding 
increases due to smaller kinetic energies of the heavy quarks. This implies into smaller $P-\rho$
overlap and thereby a smaller coupling. For the $NBP$, Fig.~\ref{fig2} shows that the $B-P$ overlap 
increases because the large-$r$ part of the light quark distribution in $B$ is cut off by the 
one from $P$, which explains the increased values of the couplings for heavier baryons. Fig.~\ref{fig2}
makes the physics transparent and explains the modest effects~on~the~couplings. 

\vspace{-1.0cm}

\vspace{-0.5cm}
We have also computed the coupling constants $g_{PP\rho}$ and $g_{NBP}$ of the 
Lagrangians in~\cite{Haidenbauer:2007jq} by matching the $^3{\rm P}_0$ transition
amplitude ${\cal M}_{h_1h_2h_3}$ in (\ref{trans}) to the one calculated with 
those Lagrangians. The matching is done at tree level at $\bq^2=0$~\cite{3P0,Downum:2006re}. 
Taking the typical values for $\gamma$ of the literature, $\gamma = 0.4-0.5$~\cite{Downum:2006re}, 
the matching leads to $g_{\pi\pi\rho } = 5.85-7.32$ and $g_{NN\pi} = 10.83-13.54$, that are in very 
good agreement with phenomenology, and $g_{KK\rho } = 2.79-3.49$, $g_{DD\rho } = 2.34-2.90$, 
$g_{N\Lambda K} = 12.65-15.81$, $g_{N\Lambda_c D} = 13.56-16.95$. In Tab.~\ref{tab2} we 
collected the ratios of these couplings and quoted results from the literature. The ratios include 
isospin factors, as in (\ref{mes-SU4}) and (\ref{bar-SU4}){\textemdash}for exact SU(4) symmetry, 
the ratios are~1. The value for $g_{DD\rho }$ agrees well with VMD~\cite{{Mat98},{Lin00a}}, 
QCDSR~\cite{Bracco:2011pg}, and lattice QCD~\cite{Can:2012tx}, and agrees within a factor of~2 
with DS-BS~\cite{ElBennich:2011py} and holographic QCD~\cite{Ballon-Bayona:2017bwk}. 

\vspace{-0.25cm}
\begin{table}[h]
\caption{Ratios of three-hadron couplings{\textemdash}for exact SU(4) symmetry, all ratios should be equal 
to~1 (see text). }
\label{tab2}  
\begin{center}
\begin{tabular}{lccc}
\hline\noalign{\smallskip}
\;\;\;\;$P/P'$                       & $\pi/K$ & $\pi/D$ & $K/D$  \\
$^3{\rm P}_0\; {\cal R}(0)$          & 1.05    & 1.26    & 1.19   \\ 
$^3{\rm P}_0\;{\cal R}(- m^2_\rho)$  & 0.99    & 1.07    & 1.08   \\
Ref.~\cite{ElBennich:2011py}         & 1.09    & 0.21    & 0.19   \\
Ref.~\cite{Ballon-Bayona:2017bwk}    & 1.11    & 2.23    & 2.00   \\
\noalign{\smallskip}\hline\noalign{\smallskip}
$BP/B'P'$                            & $N\pi/\Lambda K$ & $N\pi/\Lambda_cD$ & $\Lambda K/\Lambda_cD$ \\
$^3{\rm P}_0\; {\cal R}(0)$          & 0.89 & 0.83 & 0.92 \\
Ref.~\cite{lc-sr}                    & --- & --- & 0.68 \\
\noalign{\smallskip}\hline
\end{tabular}
\end{center}
\end{table}

\vspace{-0.5cm}
\noindent{\it Summary.} 
We used a $^3{\rm P}_0$ quark-pair creation model with nonrelativistic quark-model wave functions
to investigate the effects of SU(4) symmetry breaking in the $DD\rho$ and $N\Lambda_c D$ couplings, 
the most relevant for the $\bar DN$ and $DN$ interactions~\cite{{Haidenbauer:2007jq},{Haidenbauer:2010ch}}. 
The quark masses in the Hamiltonian~(\ref{QM-H}) are the only source of SU(4) breaking. The predictions of the model 
are reliable for low-momentum transfers in the vertices. The pattern found for SU(4) breaking 
for momenta $\bq^2 \approx 0$ in the $PP\rho$ amplitudes is $A_{DD\rho} < A_{KK\rho} 
< A_{\pi\pi\rho}$, while in the $NBP$ it is $A_{N\Lambda_c D} > A_{N\Lambda K} > A_{NN\pi}$.  
Since the $DD\rho$ (and $DD\omega$) coupling is more important for the $\bar DN$ cross section 
than the $N\Lambda_c D$ (and $N\Sigma_c D$) coupling, at least in the calculations 
in~\cite{{Haidenbauer:2007jq},{Haidenbauer:2008ff},{Haidenbauer:2010ch},{Fontoura:2012mz}}, 
our results indicate that the use of SU(4) symmetry for the coupling constants could be a 
reasonable first approximation, in line with other studies in the literature 
\cite{Mat98,Lin00a,Can:2012tx,lc-sr,{Ballon-Bayona:2017bwk}}.
Clearly, for estimating the impact of our findings 
for the SU(4) breaking on $DN$ cross sections, and also binding energies of $D$-mesic nuclei, 
further detailed studies are required. Finally, we note that the symmetry breaking pattern we 
found for $PP\rho$ couplings is opposite to that in Ref.~\cite{ElBennich:2011py}, but it agrees
with the one in the holographic QCD calculation in~\cite{Ballon-Bayona:2017bwk}. We found also
an opposite ratio for ${N\Lambda K}/{N\Lambda_c D}$ to the one in~\cite{lc-sr}. 
Further studies are needed for full clarification.

\vskip 0.2cm
Work partially supported by the Brazilian agencies CNPq, Grants No. 305894/2009-9 (GK) and 150659/2015-6 (CEF)
and FAPESP Grant No. 2013/01907-0 (G.K.).
%
%

\end{document}